\documentclass[aps,showpacs,nofootinbib,twocolumn]{revtex4}
\usepackage{graphicx}
\usepackage{color}
\usepackage[T1]{fontenc}
\usepackage[utf8]{inputenc}
\usepackage{amssymb}
\usepackage{amsthm}
\usepackage{bbm}
\usepackage{mathtools}
\usepackage{float}
\usepackage[hidelinks]{hyperref}
\usepackage{subfigure}

\usepackage[a4paper,margin=2cm,footskip=2cm]{geometry}
\begin{document}
    % Title
    \title{Stronger Schr\"odinger-like Uncertainty Relations}
    \author{Qiu-Cheng Song}
    \email{songqiucheng12@mails.ucas.ac.cn}
\author{Cong-Feng Qiao$^{1,2}$\footnote{Corresponding author, qiaocf@ucas.ac.cn}}
\affiliation{School of Physics, University of Chinese Academy of Sciences, YuQuan Road 19A, Beijing 100049, China\\
$^2$CAS Center for Excellence in Particle Physics, Beijing 100049, China}

\begin{abstract}
Uncertainty relation is one of the fundamental building blocks of quantum theory. Nevertheless, the traditional uncertainty relations do not fully capture the concept of incompatible observables. Here we present a stronger Schr\"odinger-like uncertainty relation, which is stronger than the relation recently derived by L. Maccone and A. K. Pati [Phys. Rev. Lett. 113 (2014) 260401]. Furthermore, we give an additive uncertainty relation which holds for three incompatible observables, which is stronger than the relation newly obtained by S. Kechrimparis and S. Weigert [Phys. Rev. A 90 (2014) 062118] and the simple extension of the Schr\"odinger uncertainty relation.
\end{abstract}

    \maketitle

% % % Introduction % % %

\section{Introduction}
Uncertainty is one of the distinct features of quantum theory. The concept of uncertainty principle was first introduced by Heisenberg \cite{heis}. The original form of uncertainty relation was derived by Kennard \cite{Kennard} and Weyl \cite{Weyl}. Indeed, the uncertainty relation is a mathematical description of trade-off relation in the measurement statistics of two incompatible observables. It refers to the preparation of the system which has intrinsic spreads in the measurement outcomes for independent measurements. Notably, it does not mean that two incompatible observables are impossible to be measured simultaneously on a quantum system \cite{Peres}.
The best known formula of uncertainty relation is the Heisenberg-Robertson uncertainty relation, which bounds the product of a pair of variances through the expectation value of their commutator \cite{Robertson}. It reads
\begin{eqnarray}\label{ineqr1}
\Delta A^2\Delta B^2\geq
\left|\frac{1}{2}\langle\psi|[A,B]|\psi\rangle\right|^2,
\end{eqnarray}
for arbitrary observables $A$, $B$, and any state $|\psi\rangle$, where the
variances of an observable $X$ in state $|\psi\rangle$ is defined as
$\Delta X^2=\langle\psi|X^2|\psi\rangle-\langle\psi|X|\psi\rangle^2$ and
the commutator is defined by $[A,B]=AB-BA$. A stronger extension of the uncertainty relation (\ref{ineqr1}) was made by Schr\"odinger \cite{schroedinger}, namely
\begin{eqnarray}\label{ineqr2}
\Delta A^2\Delta B^2\geq\left|\frac 12 \langle[A,B]\rangle\right|^2
+\left|\frac{1}{2}\langle\{A,B\}\rangle - \langle A\rangle\langle B\rangle\right|^2,
\end{eqnarray}
where the anti-commutator is defined by $\{A,B\}=AB+BA$, and $\langle X\rangle$ denotes the expectation value of $X$.

Uncertainty relations are significant in physics, e.g. quantum mechanics and quantum information \cite{PBusch,HHofmann,OGuhne,CAFuchs}. Traditionally the uncertainty relations try to quantitatively express the impossibility of joint sharp preparation of incompatible observables. However, in practice, they do not always capture the notion of incompatible observables since they become trivial in some cases. Recently, Maccone and Pati derived two stronger uncertainty relations based on the sum of $\Delta A^2$ and  $\Delta B^2$ \cite{mp}, which to a large extent can avoid the triviality problem and provide more stringent bounds for incompatible observables on the quantum state. The first inequality is
\begin{eqnarray}\label{ineqr3}
\Delta A^2 + \Delta B^2\geq
\pm i\langle[A,B]\rangle+|\langle\psi|A \pm iB|\psi^\perp\rangle|^2,
\end{eqnarray}
where $|\psi^\perp\rangle$ is an arbitrary state orthogonal to the state $|\psi\rangle$,  the sign on the right-hand side of the inequality takes $+(-)$ while $ i\langle[A,B]\rangle$ is positive (negative). The second inequality is
\begin{eqnarray}\label{ineqr4}
\Delta A^2 + \Delta B^2\geq
\frac 12|\langle\psi^\perp_{A+B}|A+B|\psi\rangle|^2,
\end{eqnarray}
where $|\psi^\perp_{A+B}\rangle\propto(A+B-\langle A + B\rangle)|\psi\rangle$ is a state orthogonal to $|\psi\rangle$. Maccone and Pati also derived an amended Heisenberg-Robertson uncertainty relation, i.e.
\begin{eqnarray}\label{ineqr6}
\Delta A\Delta B\geq\frac{\pm i\frac 12\langle[A,B]\rangle}
{1-\frac 12|\langle\psi|\frac{A}{\Delta A}\pm i\frac {B}{\Delta B}|\psi^\perp\rangle|^2}\ ,
\end{eqnarray}
which is stronger than the Heisenberg-Robertson uncertainty relation.

Two noncommutative sharp observables as well as three pairwise noncommutative sharp observables are incompatible, whatever the state of the system might be. Recently, two Heisenberg uncertainty relations for three canonical observables were obtained by Kechrimparis and Weigert \cite{Kechrimparis}. The multiplicative uncertainty relation reads
\begin{eqnarray}\label{ineqr7}
\Delta p\Delta q\Delta r\geq(\frac{\hbar}{\sqrt{3}})^{\frac{3}{2}}\ ,
\end{eqnarray}
where the Schr\"{o}dinger triple $(p,q,r)$ satisfies the commutation relations
\begin{eqnarray}\label{ineqr8}
[q,p]=[p,r]=[r,q] = i \hbar\ .
\end{eqnarray}
Here, the observable $r=-q-p$.
They also gave an additive uncertainty relation for the Schr\"{o}dinger triple $(q,p,r)$, it reads
\begin{eqnarray}\label{ineqr9}
\Delta p^2+\Delta q^2+\Delta r^2\geq \sqrt{3} \hbar\ .
\end{eqnarray}

In this work, two new Schr\"odinger-like uncertainty relations for the sum and product of variances of two observables by extending the Schr\"odinger uncertainty relation (\ref{ineqr2}) are obtained. An uncertainty relation for three observables will be given, which is stronger than the uncertainty relation given by Kechrimparis and Weigert, and we will exhibit its property in case of spin-1 system.
\section{ Schr\"odinger-like Uncertainty Relation}
\begin{figure}[hbt]
\begin{center}
%\subfigure{\includegraphics[width=0.33\textwidth]{aa.pdf}}
%\subfigure{\includegraphics[width=0.33\textwidth]{bb.pdf}}
%\subfigure{\includegraphics[width=0.33\textwidth]{cc.pdf}}
\includegraphics[width=0.31\textwidth]{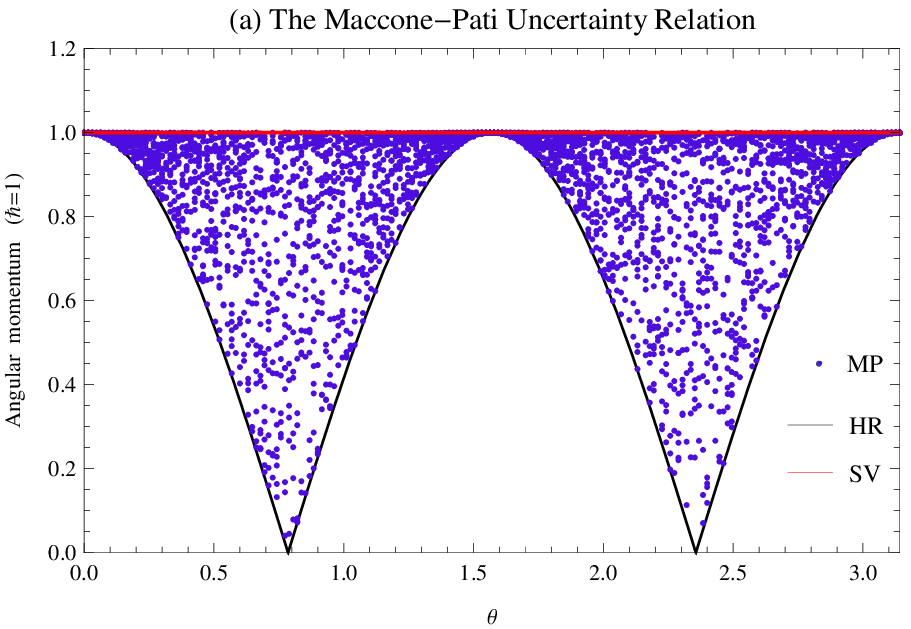}
\mbox{\hspace{0.1cm}}
\includegraphics[width=0.31\textwidth]{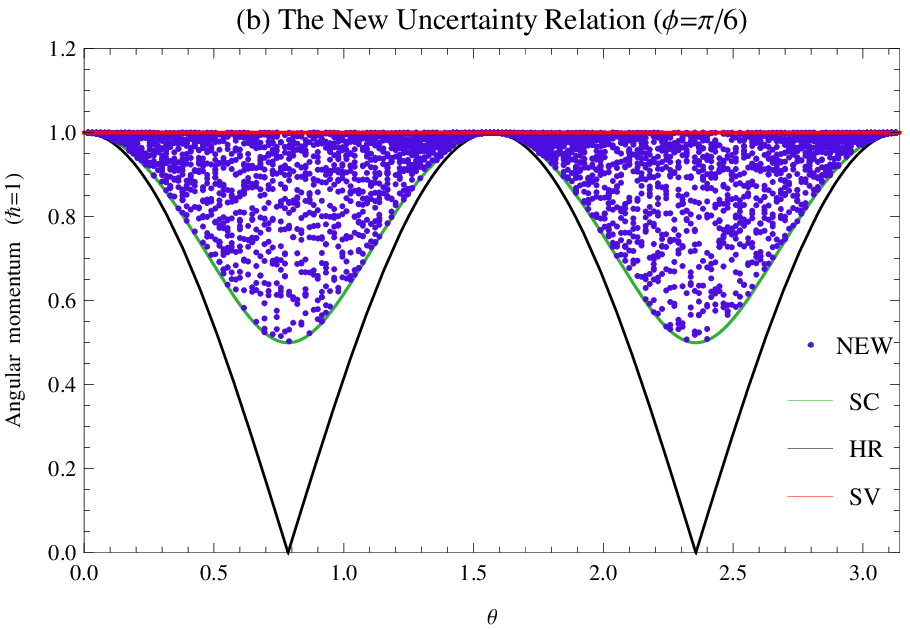}
\mbox{\hspace{0.1cm}}
\includegraphics[width=0.31\textwidth]{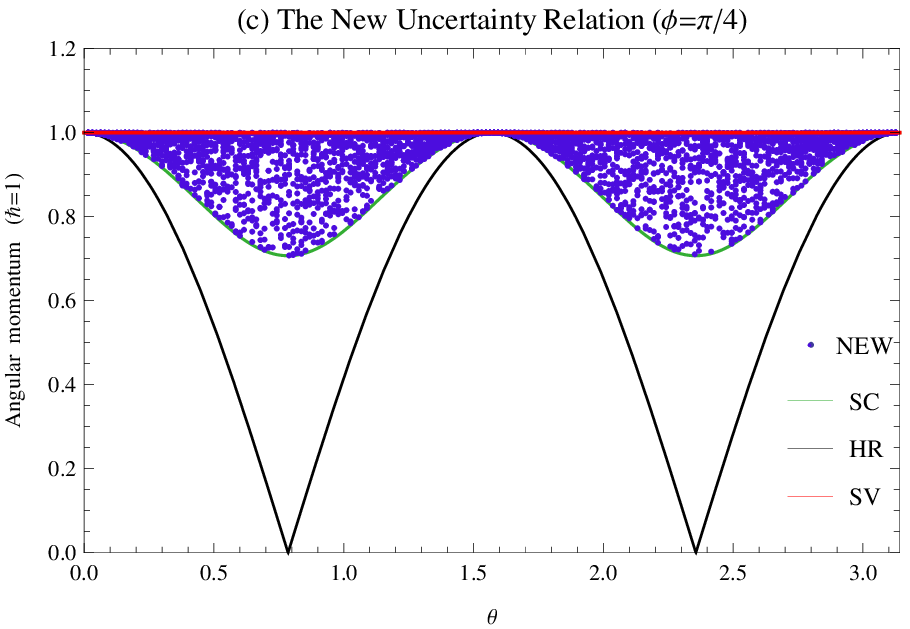}
\end{center}
\vspace{-.6cm}
\caption{Example of comparison between the Maccone-Pati uncertainty relation (MP) (\ref{ineqr3}) and the new uncertainty relation (NEW) (\ref{ineqa}). Note that the new uncertainty relation (\ref{ineqa}) is stronger than the relation (\ref{ineqr3}).  We choose two components of the angular momentum $A=J_x$ and $B=J_y$  for a spin-1 particle, and a family of states parameterized by $\theta$ and $\phi$ as  $|\psi\rangle=\cos\theta|1\rangle+\sin\theta e^{i\phi}|-1\rangle$, with $|\pm1\rangle$ being eigenstates of $J_z$ corresponding to the eigenvalues of $\pm1$. The upper red line denotes the sum of variances $\Delta J_x^2+\Delta J_y^2$ (SV). The blue points exhibit domains of (\ref{ineqr3}) in ({\bf a}) and (\ref{ineqa}) in ({\bf b}: $\phi=\pi/6$) and ({\bf c: $\phi=\pi/4$}) with 20 randomly chosen states $|\psi^\perp\rangle$ for each of the 200 values of the phase $\theta$. The green curve is the lower bound given by the Schr\"odinger uncertainty relation (SC) (\ref{ineqc}). The black curve is the lower bound set by the Heisenberg-Robertson uncertainty relation (HR) (\ref{ineqr1}). The relation (\ref{ineqr3}) gives the same results for any value of $\phi$ ({\bf a}). If $\phi$ is not equal to $0$ and $\pi$, the new uncertainty relation (\ref{ineqa}) always give nontrivial bound ({\bf b}) and ({\bf c}). When $\phi$ is equal to $0$ or $\pi$, the relation (\ref{ineqa}) reduces to the relation (\ref{ineqr3}) and have the same results as ({\bf a}).}\label{two}
\end{figure}

The first Schr\"odinger-like uncertainty relation reads
\begin{align}\label{ineqa}
\Delta A^2 + \Delta B^2\geq&
|\langle[A,B]\rangle + \langle\{A,B\}\rangle-2\langle A\rangle\langle B\rangle|\notag\\
&+|\langle\psi|A-e^{i\alpha}B|\psi^\perp\rangle|^2\ ,
\end{align}
which is valid for arbitrary states $|\psi^\perp\rangle$ orthogonal to the state of the system $|\psi\rangle$ and stronger than Maccone and Pati's uncertainty relation (\ref{ineqr3}) (Fig. \ref{two}), where $\alpha$ is a real constant. If $\langle \{A,B\} \rangle-2\langle A\rangle\langle B\rangle >0$, then  $\alpha = \arctan \tfrac{-i\langle[A,B]\rangle} {\langle\{A,B\} \rangle- 2\langle A\rangle\langle B\rangle}$; if $\langle\{A,B\}\rangle-2\langle A\rangle \langle B\rangle<0$, then $\alpha = \pi+\arctan \tfrac{-i\langle[A,B] \rangle} {\langle\{A,B\}\rangle-2\langle A\rangle\langle B\rangle}$; and while
$\langle\{A,B\} \rangle-2\langle A\rangle\langle B\rangle=0$, it reduces to (\ref{ineqr3}). Removing the last term of (\ref{ineqa}), it then turns into
\begin{eqnarray}\label{ineqc}
\Delta A^2 + \Delta B^2\geq
|\langle[A,B]\rangle+\langle\{A,B\}\rangle-2\langle A\rangle\langle B\rangle|\ ,
\end{eqnarray}
which is implied by the Schr\"odinger uncertainty relation (\ref{ineqr2}).

The second Schr\"odinger-like uncertainty relation is
\begin{eqnarray}\label{ineqb}
\Delta A^2\Delta B^2\geq\frac{\left|\tfrac 12\langle[A,B]\rangle\right|^2
+ \left|\tfrac{1}{2}\langle\{A,B\}\rangle-\langle A\rangle\langle B\rangle\right|^2}
{(1-\tfrac 12|\langle\psi|\tfrac{A} {\Delta A}-e^{i\alpha}\tfrac {B}{\Delta B}|\psi^\perp\rangle|^2)^2},
\end{eqnarray}
which is stronger than the Schr\"odinger uncertainty relation (\ref{ineqr2}) and reduces to (\ref{ineqr6}) when $\langle\{A,B\} \rangle-2\langle A\rangle\langle B\rangle=0$.

\emph{Proof}: To prove the uncertainty relation (\ref{ineqa}),
we start by introducing a general inequality
\begin{eqnarray}
\|c_A\bar{A}|\psi\rangle
  -c_B e^{i\tau}\bar{B}|\psi\rangle
  +c(|\psi\rangle-|\phi\rangle)\|^2\geq 0\ ,
 \end{eqnarray}
with $\bar{A}=A-\langle\psi|A|\psi\rangle$, $\bar{B} = B- \langle \psi|B| \psi \rangle$; $c_A$, $c_B$, $c$ and $\tau$ being real numbers, and  $|\phi\rangle$ being an arbitrary state.
Calculating the modulus squared, we have
\begin{eqnarray}\label{ineq0}
c_A^2\Delta A^2+c_B^2\Delta B^2\geq-\lambda c^2-c_A c_B c\beta + c_A c_B\delta\ .
\end{eqnarray}
Here, $\Delta A^2$ and $\Delta B^2$ are the variances of $A$ and $B$
calculated on $|\psi\rangle$, respectively. $\lambda\equiv 2(1-\text{Re} [\langle\psi|\phi \rangle])$, $\beta\equiv 2\text{Re} [\langle\psi| (-\bar{A}/c_B+e^{-i\tau}\bar{B}/c_A| \phi\rangle]$,
and $\delta\equiv2\text{Re}[e^{i\tau}\langle\psi|\bar{A}\bar{B}|\psi\rangle]$.
Choosing the value of $c$ that maximizes the right-hand-side of (\ref{ineq0}), namely $c=-c_A c_B\beta/2\lambda$, we then get
\begin{eqnarray}\label{ineq1}
c_A^2\Delta A^2+c_B^2\Delta B^2\geq
\frac{(c_A c_B\beta)^2}{4\lambda} + c_A c_B\delta\ .
\end{eqnarray}
We can further choose $c_A=1$ and $c_B=1$, we obtain
\begin{align}\label{ineq2}
\Delta A^2 + \Delta B^2\geq&
\frac{\{\text{Re}[\langle\psi|-\bar{A}+e^{-i\tau}\bar{B}|\phi\rangle]\}^2}
 {2(1-\text{Re}[\langle\psi|\phi\rangle])}\notag\\
 &+2\text{Re}[e^{i\tau}\langle\psi|\bar{A}\bar{B}|\psi\rangle]\ .
\end{align}
Suppose $|\phi\rangle = \cos \theta| \psi\rangle + e^{i\phi} \sin\theta |\psi^{\perp}\rangle$,
where $| \psi^{\perp}\rangle$ is orthogonal to
$|\psi\rangle$. By taking the limit $\theta\rightarrow 0$, so that
 $|\phi\rangle \rightarrow |\psi\rangle$. Then
the inequality (\ref{ineq2}) yields
\begin{align}\label{ineq3}
\Delta A^2 + \Delta B^2\geq&
\{\text{Re}[e^{i\phi}\langle\psi|
-A+e^{-i\tau}B|\psi^\perp\rangle]\}^2\notag\\
&+2\text{Re}[e^{i\tau}\langle\psi|\bar{A}\bar{B}|\psi\rangle]\ .
\end{align}
There exists $\tau=-\alpha$, so that $e^{i\tau}\langle\psi|\bar{A}\bar{B}|\psi\rangle$ is real
and can be written as $|\langle\bar{A}\bar{B}\rangle|$, then the second term of (\ref{ineq3}) becomes $\{\text{Re}[e^{i\phi}\langle\psi|
-A+e^{i\alpha}B|\psi^\perp\rangle]\}^2$. Choosing a proper phase $\phi$ which makes the term in the square brackets to be real,
it can then be expressed as $|\langle\psi|A-e^{i\alpha}B|\psi^\perp\rangle|^2 $.
In the end, the inequality (\ref{ineq3}) turns to
\begin{align}\label{ineq4}
\Delta A^2 + \Delta B^2\geq
2|\langle\bar{A}\bar{B}\rangle|+|\langle \psi|A-e^{i\alpha}B|\psi^\perp\rangle|^2\ .
\end{align}
Of the quantity $|\langle\bar{A}\bar{B}\rangle|$, it is easy to see that
\begin{align}\label{ineq5}
\langle\bar{A}\bar{B}\rangle=&\frac{1}{2} \langle[\bar{A},
\bar{B}]\rangle+\frac{1}{2}\langle\{\bar{A}, \bar{B}\}\rangle\notag\\
=&\frac{1}{2}\langle[A,B]\rangle+\frac{1}{2} \langle\{A,B\}\rangle
-\langle A\rangle\langle B\rangle\ .
\end{align}
Substituting $\langle\bar{A}\bar{B}\rangle$ in (\ref{ineq4}) into (\ref{ineq5}), one obtains the uncertainty relation (\ref{ineqa}).

To prove the improved Schr\"odinger uncertainty relation (\ref{ineqb}),
we can choose $m=\Delta B$ and $n=\Delta A$ in (\ref{ineq1}), which then becomes
\begin{eqnarray}\label{ineq6}
\Delta A\Delta B \geq&\frac{\Delta A \Delta B\{\text{Re}[\langle\psi|
-\frac{\bar{A}}{\Delta A}+e^{-i\tau}\frac{\bar{B}}{\Delta B}|\phi\rangle]\}^2}
 {4(1-\text{Re}[\langle\psi|\phi\rangle])}\notag\\
 &+\text{Re}[e^{i\tau}\langle\psi|\bar{A}\bar{B}|\psi\rangle]\ .
\end{eqnarray}
Taking the $|\phi\rangle\rightarrow|\psi\rangle$ limit and
using the same procedure described above, the inequality (\ref{ineq6}) becomes
\begin{eqnarray}\label{ineq7}
&\Delta A\Delta B\geq\frac{\Delta A\Delta B}{2}\left|\langle\psi|\frac{A}{\Delta A}-e^{i\alpha}\frac {B}{\Delta B}|\psi^\perp\rangle\right|^2\notag\\
 &+\left|\frac 12\langle[A,B]\rangle+\frac 12\langle\{A,B\}\rangle-\langle A\rangle\langle B\rangle\right|,
\end{eqnarray}
which tells
\begin{eqnarray}\label{ineq8}
\Delta A\Delta B\geq\frac{\left| \frac 12\langle[A,B]\rangle
+\frac 12\langle\{A,B\}\rangle-\langle A \rangle \langle B\rangle\right|}
{(1-\frac 12|\langle\psi\left| \frac{A}{\Delta A}-e^{i\alpha}\frac {B}{\Delta B}|\psi^\perp\rangle\right|^2)}\ .
\end{eqnarray}
From (\ref{ineq8}) one can simply obtain the improved Schr\"odinger-like uncertainty relation (\ref{ineqb}).

Note that as this work was finished, there appeared several papers relating to Maccone and Pati's work \cite{mp}. Eq. (4) of Ref. \cite{ba} and Eq. (55) of Ref. \cite{sun} are similar to our uncertainty relation (\ref{ineqa}).  Ref. \cite{ba} mentions that Eq.  (3) of Ref. \cite{mp} may still experience triviality problem in special case when $|\psi^\perp\rangle =\frac{(A-\langle A\rangle)|\psi\rangle}{\Delta A}$ or $|\psi^\perp\rangle =\frac{(B-\langle B\rangle)|\psi\rangle}{\Delta B}$, which means the uncertainty relations in this work also have such drawback. In practice, if $|\psi^\perp\rangle$ is chosen properly, one can certainly get rid of such triviality problem, e.g, taking the $|\psi^\perp\rangle$ to be orthogonal to $|\psi\rangle$ but not orthogonal to $(A-e^{-i\alpha}B)|\psi\rangle$. Moreover, it is worth mentioning that certain kinds of uncertainty relations, e.g. \cite{li,huang}, are quantum state independent and hence immune from the triviality problem.

\section{Uncertainty Relation for Three Observables}
\subsection{New uncertainty relation}

One may generalize the Schr\"odinger uncertainty relation (\ref{ineqr2}) to three observables trivially, that is
\begin{align}\label{ineqsch}
&\Delta A^2 + \Delta B^2+\Delta C^2\geq\notag\\
&\frac 12\{\left|\langle[A,B]\rangle+\langle\{A,B\}\rangle-2\langle A\rangle\langle B\rangle\right|\notag\\
&+\left|\langle[B,C]\rangle+\langle\{B,C\}\rangle-2\langle B\rangle\langle C\rangle\right|\notag\\
&+\left|\langle[C,A]\rangle+\langle\{C,A\}\rangle-2\langle C\rangle\langle A\rangle\right|\}\ ,
\end{align}
which is simply the sum of the inequality (\ref{ineqc}). However, we will prove that the following more stringent inequality exists:
\begin{align}\label{th1}
&\Delta A^2+\Delta B^2+\Delta C^2\geq
\frac{1}{3}\left|\langle\psi^\perp_{ABC}|A+B+C|\psi\rangle\right|^2 \notag\\
&+\frac{\sqrt3}{3}\left|i\langle[A,B,C]\rangle\right|
+\frac{2}{3}\left|\langle\psi|A+e^{\pm i \frac{2\pi}{3}}B
+e^{\pm i \frac{4\pi}{3}}C|\psi^\perp\rangle\right|^2,
\end{align}
which is valid for arbitrary states $|\psi^\perp\rangle$ orthogonal to the state of the system $|\psi\rangle$, where $|\psi^\perp_{ABC} \rangle \propto(A+B+C-\langle A+B+C\rangle)|\psi\rangle$, $\langle[A,B,C]\rangle\equiv\langle[A,B]\rangle+ \langle[B,C]\rangle+\langle[C,A]\rangle$, the sign in the last term of (\ref{th1}) is $+(-)$ when $ i\langle[A,B,C]\rangle$ is positive (negative).

Applying Schr\"odinger triple $(p,q,r)$ to uncertainty relation (\ref{th1}), the Kechrimparis and Weigert's relation (\ref{ineqr9}) can be readily obtained. Choosing an arbitrary state $|\psi\rangle$ and letting $A=q$, $B=p$ and $C=r$, the uncertainty relation (\ref{th1}) then goes like
\begin{align}\label{tha}
&\Delta q^2+\Delta p^2+\Delta r^2\geq\sqrt{3}\hbar\notag\\
&+\frac{2}{3}\left|\langle\psi|q+e^{\pm i \frac{2\pi}{3}}p
+e^{\pm i \frac{4\pi}{3}}r|\psi^\perp\rangle\right|^2.
\end{align}
Discarding the last term on the right-hand side of the inequality, one obtains the Kechrimparis and Weigert's relation (\ref{ineqr9}).
Furthermore, the uncertainty relation (\ref{th1}) is obviously stronger than the Kechrimparis and Weigert's relation (\ref{ineqr9}), since the extra term in (\ref{tha}) is nonnegative.

As suggested to us by an anonymous Referee, the most uncertainty relations do not depend on the order that one chooses to label the operators, but the  three terms on the right hand side of inequality (\ref{th1}) do not remain invariant when one changes in the order of the three operators or under sign flips (eg sunbstituting $A$ with $-A$). To get an symmetrical uncertainty relation for three incompatible observables, as we all know the equation
\begin{align}\label{thb}
\Delta A^2+\Delta B^2+\Delta C^2=\Delta (\pm A)^2+\Delta (\pm B)^2+\Delta (\pm C)^2,
\end{align}
 when we use inequality (\ref{th1}), the right hand side of the equality (\ref{thb}) have four different lower bounds ${\cal L}_{i} (i=1,2,3,4)$ when we choose the same $|\psi^\perp\rangle$ in general,  there exist one ${\cal L}_{i}$ that has the term $|\langle [A,B,C]\rangle|=|\langle [A,B]\rangle| + |\langle [B,C]\rangle| + |\langle[C,A]\rangle|$. We must have the inequality
\begin{align}\label{thc}
&\Delta A^2+\Delta B^2+\Delta C^2\notag\\
\geq&\frac{\sqrt3}{3}\{|\langle [A,B]\rangle| + |\langle [B,C]\rangle| + |\langle[C,A]\rangle|\}
\end{align}
which do not depend on the order that one chooses to label the operators and works as a suitable generalisation of the Heisenberg-Robertson inequality (\ref{ineqr1}). The relation (\ref{ineqr9}) also can be derived from the relaion (\ref{thc}).

\emph{Proof}: To prove the uncertainty relation (\ref{th1}), we start by introducing a general inequality
\begin{align}\label{th}
\|\bar{A}|\psi\rangle
  + e^{i\rho}\bar{B}|\psi\rangle+ e^{i\sigma}\bar{C}|\psi\rangle
  +c(|\psi\rangle-|\phi\rangle)\|^2 \geq 0,
 \end{align}
with $\bar{A}=A-\langle\psi|A|\psi\rangle$, $\bar{B}=B-\langle\psi|B|\psi\rangle$,
$\bar{C}=C-\langle\psi|C|\psi\rangle$ and $\rho$, $\sigma$, $c$ real constants, respectively, where $|\phi\rangle$ is an arbitrary state. This inequality has good results when $\rho$ and $\sigma$ are equal to $\pm\frac{2\pi}{3}$ and $\pm\frac{4\pi}{3}$ respectively (see Appendix). Simplifying the modulus squared, we find
\begin{align}\label{th2}
\Delta A^2+\Delta B^2+\Delta C^2\geq-\lambda c^2+\beta c-\delta,
\end{align}
by defining
\begin{align}
\beta=&2\text{Re}(\langle\psi|\bar{A}+e^{\mp i\frac{2\pi}{3}}\bar{B}
+e^{\mp i\frac{4\pi}{3}}\bar{C}|\phi\rangle),\\
\lambda=&2(1-\text{Re}(\langle\psi|\phi\rangle)),\\
\delta=&-\frac12\langle\{A,B,C\}\rangle\pm i\frac{\sqrt{3}}{2}\langle[A,B,C]\rangle\notag\\
 &+\langle A\rangle\langle B\rangle+\langle A\rangle\langle C\rangle+\langle B\rangle\langle C\rangle\ .\label{pi}
\end{align}
Here, $\langle\{A,B,C\}\rangle\equiv\langle\{A,B\}\rangle + \langle\{A,C\}\rangle+\langle\{B,C\}\rangle$.
Noticing
\begin{align}\label{th6}
&\Delta(A+B+C)^2\notag\\
=&\Delta A^2 + \Delta B^2+ \Delta C^2+\langle\{A,B,C\}\rangle\notag\\
 &-2(\langle A\rangle\langle B\rangle+\langle A\rangle\langle C\rangle+\langle B\rangle\langle C\rangle)\ ,
\end{align}
the equality (\ref{pi}) can then be reexpressed as
\begin{align}\label{th7}
\delta=&-\frac12[\Delta(A+B+C)^2-(\Delta A^2+\Delta B^2+\Delta C^2)]\notag\\
&\pm i\frac{\sqrt{3}}{2}\langle[A,B,C]\rangle\ .
\end{align}
Assuming $|\phi\rangle =\cos\theta|\psi\rangle +e^{i\phi}\sin\theta|\psi^{\perp}\rangle$ and
using the same techniques employed in deriving (\ref{ineqa}), we obtain
\begin{align}
&\Delta A^2 + \Delta B^2+ \Delta C^2\geq
\frac{1}{3}\Delta (A+B+C)^2\notag\\
&\mp i\frac{\sqrt3}{3}\langle[A,B,C]\rangle
+\frac{2}{3}\left|\langle\psi|A+e^{\mp i \frac{2\pi}{3}}B+
e^{\mp i \frac{4\pi}{3}}C|\psi^\perp\rangle\right|^2\ ,
\end{align}
which is equivalent to the uncertainty relation (\ref{th1}) since $\Delta (A+B+C)^2=\left|\langle\psi^\perp_{ABC}|A+B+C|\psi\rangle\right|^2$. Here the sign should be chosen properly so that
$\mp i\frac{\sqrt3}{3}\langle[A,B,C]\rangle$ (a real quantity) is positive.

Recently, Ref.\cite{sun} gave the variance-based uncertainty equalities for any pairs of incompatible observables $A$ and $B$ . When applied to three observables, the uncertainty equality reads
\begin{align}\label{threeequ}
&\Delta A^2+\Delta B^2+\Delta C^2\notag\\
=&\frac{1}{3}\left|\langle\psi^\perp_{ABC}|A+B+C|\psi\rangle\right|^2
+\frac{\sqrt3}{3}\left|i\langle[A,B,C]\rangle\right|\notag\\
&+\frac{2}{3}\sum^{d-1}_{n=1}\left|\langle\psi|A+e^{\pm i \frac{2\pi}{3}}B
+e^{\pm i \frac{4\pi}{3}}C|\psi^\perp_n\rangle\right|^2,
\end{align}
where $\{|\psi\rangle, {|\psi_n^{\perp}\rangle}_{n=1}^{d-1}\}$ form an orthonormal and complete basis in $d$-dimensional Hilbert space, the sign in the last term of (\ref{threeequ}) is $+(-)$ when $ i\langle[A,B,C]\rangle$ is positive (negative). If we retain only one term associated with $|\psi^{\perp}\rangle\in\{|\psi^\perp_{n}\rangle_{n=1}^{d-1}\}$ in the summation and
discard others, it reduces to the uncertainty inequality (\ref{th1}).

\begin{figure}[hbt]
\begin{center}
\subfigure{\includegraphics[width=0.33\textwidth]{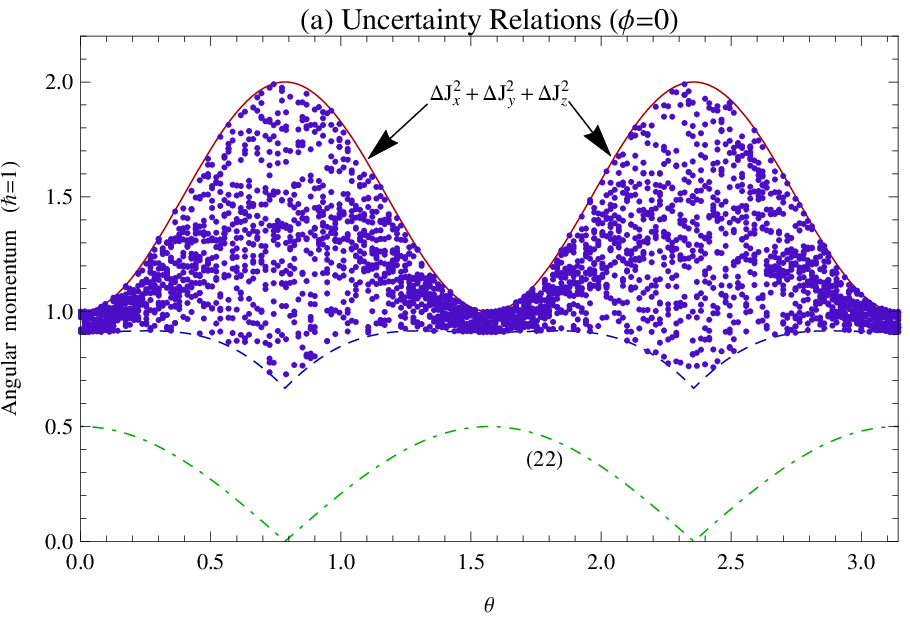}}
%\mbox{\hspace{1.0cm}}
\subfigure{\includegraphics[width=0.33\textwidth]{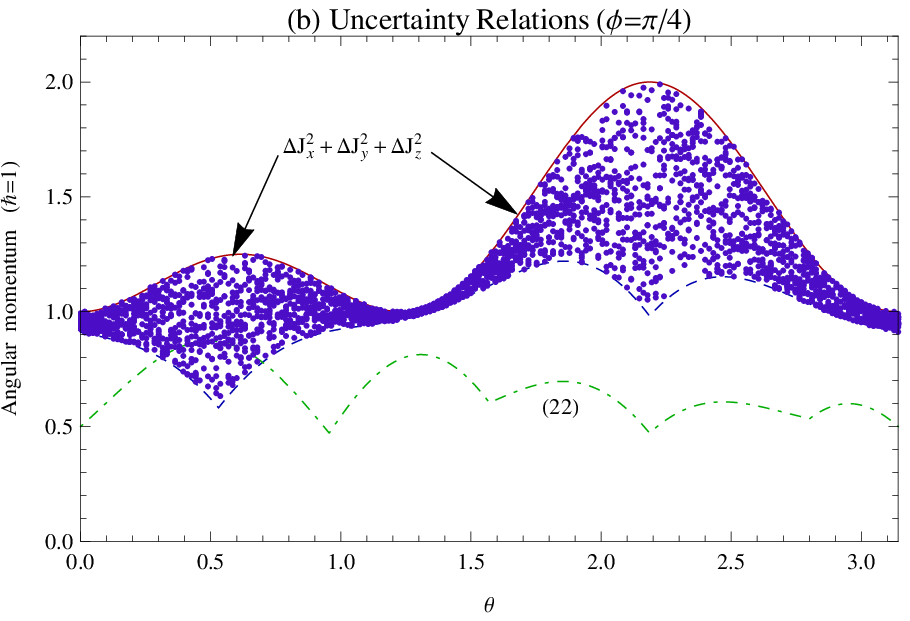}}
\end{center}
\vspace{-.5cm}
\caption{Uncertainty relation for three components of the angular momentum $A=J_x$, $B=J_y$ and $C=J_z$ for a spin-1 particle.
The state chosen is parameterized by $\theta$ and $\phi$ as $|\psi\rangle=\sin\theta\cos \phi|1\rangle+\sin\theta \sin\phi|0\rangle+\cos\theta|-1\rangle$. Here, $|\pm1\rangle$ and $|0\rangle$ are eigenstates of $J_z$ corresponding to eigenvalues $\pm1$ and $0$. The diagrams illustrate how different uncertainty relations (\ref{ineqsch}) and (\ref{th1}) restrict the possible values of the sum of variances in different values of $\phi$ ($\phi=0$ in ({\bf a}) and $\phi=\pi/4$ in ({\bf b})). The upper red curve shows $\Delta J_x^2+\Delta J_y^2+\Delta J_z^2$. The blue points exhibit domains of (\ref{th1}) with 15 randomly chosen states $|\psi^\perp\rangle$ for each of the 200 values of the phase $\theta$. The dash-dotted green curve is the bound given by the trivially generalized Schr\"odinger uncertainty relation (\ref{ineqsch}) for three observables.}\label{three}
\end{figure}

\subsection{Application to spin-1 particle state}

As an illustration of the uncertainty relation (\ref{th1}), we consider a simple case of spin-1 particle state. Let $A=J_x$, $B=J_y$ and $C=J_z$ to be three components of the angular momentum, and take
\begin{eqnarray}
|\psi\rangle=\sin\theta\cos \phi|1\rangle+\sin\theta \sin\phi|0\rangle+\cos\theta|-1\rangle\
\end{eqnarray}
and
\begin{align}
|\psi^\perp\rangle=&(\cos\theta\cos\phi\cos\beta e^{i\gamma} -\sin\phi\sin\beta)|1\rangle\notag\\
&+(\cos\theta\sin\phi\cos\beta e^{i\gamma}+\cos\phi\sin\beta )|0\rangle\notag\\
&-(\sin\theta\cos\beta e^{i\gamma})|-1\rangle\
\end{align}
as the states of system, with $|\pm1\rangle$ and $|0\rangle$ the eigenstates of $J_z$ corresponding to eigenvalues of $\pm1$ and $0$. The $\beta$ and $\gamma$ in the orthogonal state are free parameters.

In Fig. \ref{three}, we compare numerically the uncertainty relation obtained in this work, the relation (\ref{th1}), with the simply generalized Schr\"odinger uncertainty relation (\ref{ineqsch}) for three observables.

When $\phi=0$, the relation (\ref{ineqsch}) changes to
\begin{eqnarray}
\frac{1}{2}(3-\cos(4\theta))\geq \frac{1}{2}\left|\cos(2\theta)\right|.
\end{eqnarray}
Discarding the last term in relation (\ref{th1}), it then reads
\begin{eqnarray}
\frac{1}{2}(3-\cos(4\theta))\geq\frac{1}{6}\left(2\sqrt{3} \left|\cos(2\theta)\right|-\cos(4\theta)+3\right).
\end{eqnarray}
Since state $|\psi^\perp\rangle$ in (\ref{th1}) is an arbitrary state orthogonal to $|\psi\rangle$, the blue points in Fig. \ref{three} illustrate the domain of (\ref{th1}) with 15 randomly taking states $|\psi^\perp\rangle$ for each of the 200 values
of the phase $\theta$. We find the uncertainty relation (\ref{th1}) is nontrivial for all $\theta$s and stronger than the simply generalized Schr\"odinger uncertainty relation (\ref{ineqsch}).

When $\phi=\pi/4$ and $\theta\in(0,0.3067)\bigcup(0.6991,\pi)$, the uncertainty relation (\ref{th1}) is also stronger than the generalized Schr\"odinger uncertainty relation (\ref{ineqsch}). In fact, if one chooses $|\psi^\perp\rangle$ properly, the uncertainty relation (\ref{th1}) is always stronger than (\ref{ineqsch}) for any values of $\theta$. It means that the whole incompatible nature of three observables can not be simply represented by three independent pairwise incompatible observables.

\section{Conclusions}

In this work, we have obtained a stronger Schr\"odinger-like uncertainty relation (\ref{ineqa}) based on the sum of variances of two observables, which is stronger than the uncertainty relation (\ref{ineqr3}) given by Maccone and Pati. Meanwhile, we have also developed an improved Schr\"odinger-like uncertainty relation (\ref{ineqb}) which is stronger than the Schr\"odinger uncertainty relation (\ref{ineqr2}). Furthermore, we have obtained an uncertainty relation which holds for three observables, and it is proven to be stronger than the uncertainty relation (\ref{ineqr9}) given by Kechrimparis and Weigert. Finally, as an illustration, we have taken  spin-1 particle system as an example to show that the uncertainty relation (\ref{th1}) obtained in this work is stronger than the simply generalized Schr\"odinger uncertainty relation, which means that the whole incompatible nature of three observables can not be simply represented by the natures of three independent pairwise incompatible observables.

%%%%%%%%%%%%%%%%%%%%%%%%%%%%%%%%%%%%%%%%%%%%%%%%%%%%%%%%%%%%%%%%%%%%%%
\vspace{.1cm}
{\bf Acknowledgments}
We are grateful to Junli Li and Zhiyong Bao for helpful discussions, to Rui Xu for reading through the manuscript and suggestions. This work was supported in part by the Ministry of Science and Technology of the People's Republic of China (2015CB856703), and by the National Natural Science Foundation of China(NSFC) under the grants 11175249 and 11375200.

%%%%%%%%%%%%%%%%%%%%%%%%%%%%%%%%%%%%%%%%%%%%%%%%%%%%%%%%%%%%%%%%%%%%%%
\onecolumngrid
\section*{Appendix}

To illustrate why we choose $\rho=\pm\frac{2\pi}{3}$ and $\sigma=\pm\frac{4\pi}{3}$ in the inequality (\ref{th}), we start from the following three inequalities
\begin{subequations}\label{eq:a}
\begin{align}
\|\bar{A}|\psi\rangle+e^{i\rho}\bar{B}|\psi\rangle+e^{i\sigma}\bar{C}|\psi\rangle\|^2 &\geq 0\ ,\label{eq:a1}\\
\|\bar{B}|\psi\rangle+e^{i\rho}\bar{C}|\psi\rangle+e^{i\sigma}\bar{A}|\psi\rangle\|^2 &\geq 0\ ,\label{eq:a2}\\
\|\bar{C}|\psi\rangle+e^{i\rho}\bar{A}|\psi\rangle+e^{i\sigma}\bar{B}|\psi\rangle\|^2 &\geq 0\ ,\label{eq:a3}
\end{align}
\end{subequations}
where $\rho,\sigma\in(0,2\pi)$, $\bar{A}=A-\langle\psi|A|\psi\rangle$, $\bar{B}=B-\langle\psi|B|\psi\rangle$,
$\bar{C}=C-\langle\psi|C|\psi\rangle$. By expanding the square modulus, we have
\begin{subequations}\label{eq:b}
\begin{align}
\Delta A^2+\Delta B^2+\Delta C^2
+2\text{Re}(e^{i\rho}\langle\bar{A}\bar{B}\rangle)
+2\text{Re}(e^{i\sigma}\langle\bar{A}\bar{C}\rangle)
+2\text{Re}(e^{i(\sigma-\rho)}\langle\bar{B}\bar{C}\rangle) \geq 0\ ,\label{b1}\\
\Delta A^2+\Delta B^2+\Delta C^2
+2\text{Re}(e^{i\rho}\langle\bar{B}\bar{C}\rangle)
+2\text{Re}(e^{i\sigma}\langle\bar{B}\bar{A}\rangle)
+2\text{Re}(e^{i(\sigma-\rho)}\langle\bar{C}\bar{A}\rangle) \geq 0\ ,\label{b2}\\
\Delta A^2+\Delta B^2+\Delta C^2
+2\text{Re}(e^{i\rho}\langle\bar{C}\bar{A}\rangle)
+2\text{Re}(e^{i\sigma}\langle\bar{C}\bar{B}\rangle)
+2\text{Re}(e^{i(\sigma-\rho)}\langle\bar{A}\bar{B}\rangle) \geq 0\ .\label{b3}
\end{align}
\end{subequations}
To evaluate the inequalities (\ref{eq:b}), we notice
\begin{align}\label{c}
&2\text{Re}(e^{i\rho}\langle\bar{E}\bar{F}\rangle)
=\cos\rho(\langle\{E,F\}-2\langle E\rangle\langle F\rangle)+i\sin\rho\langle[E,F]\rangle\ ,\\
&\Delta(A+B+C)^2
=\Delta A^2 + \Delta B^2+ \Delta C^2+\langle\{A,B,C\}\rangle
-2(\langle A\rangle\langle B\rangle+\langle A\rangle\langle C\rangle+\langle B\rangle\langle C\rangle)\ ,
\end{align}
where $E$ and $F$ are arbitrary observables, and we define $\langle\{A,B,C\}\rangle\equiv\langle\{A,B\}\rangle + \langle\{A,C\}\rangle+\langle\{B,C\}\rangle$.
Calculating (\ref{b1})+(\ref{b2})+(\ref{b3}), we obtain
\begin{align}\label{d}
\Delta A^2+\Delta B^2+\Delta C^2 \geq \mu\Delta(A+B+C)^2+i\nu\langle[A,B,C]\rangle\ ,
\end{align}
where we define
\begin{align}\label{e}
&\langle[A,B,C]\rangle\equiv\langle[A,B]\rangle+ \langle[B,C]\rangle+\langle[C,A]\rangle\ ,\\
&\mu=\frac{\cos{\rho}+\cos{\sigma} + \cos(\sigma-\rho)}{\cos{\rho}+\cos{\sigma}+\cos(\sigma-\rho)-3}\ ,\\
&\nu=\frac{\sin{\rho}-\sin{\sigma}+ \sin(\sigma-\rho)}{\cos{\rho}+ \cos{\sigma}+\cos(\sigma-\rho)-3}\ .
\end{align}
When $\rho=\pm\frac{2\pi}{3}$ and $\sigma=\pm\frac{4\pi}{3}$, $|\mu|$ and $|\nu|$ all have maximum values, namely
\begin{align}\label{f}
\mu(\rho &=\frac{2\pi}{3}\ ,\sigma=\frac{4\pi}{3})=\frac{1}{3}\ , \hspace{1.1cm}
\nu(\rho =\frac{2\pi}{3}\ ,\sigma=\frac{4\pi}{3})=-\frac{1}{\sqrt{3}}\ ;\\
\mu(\rho &=-\frac{2\pi}{3}\ ,\sigma=-\frac{4\pi}{3})=\frac{1}{3}\ ,\hspace{0.5cm}
\nu(\rho =-\frac{2\pi}{3}\ ,\sigma=-\frac{4\pi}{3})=\frac{1}{\sqrt{3}}\ .
\end{align}
Hence, the inequality (\ref{d}) becomes
\begin{align}\label{g}
\Delta A^2+\Delta B^2+\Delta C^2 \geq \frac{1}{3}\Delta(A+B+C)^2\pm \frac{i}{\sqrt{3}}\langle[A,B,C]\rangle\ .
\end{align}
To gain maximum value of the right-hand side of (\ref{g}), we should choose the sign properly so that $\pm\frac{i}{\sqrt{3}}\langle[A,B,C]\rangle$ (a real quantity) is positive.

\twocolumngrid

%%% Bibliography %%%
\bibliography{Uram}

\end{document}